# Projection Mapping Technologies for AR

## Daisuke iwai

Osaka University, Machikaneyamacho 1-3, Toyonaka, Osaka 5608531, Japan

Keywords: Projection mapping, augmented reality, computational display

**ABSTRACT**

*This invited talk will present recent projection mapping technologies for augmented reality. First, fundamental technologies are briefly explained, which have been proposed to overcome the technical limitations of ordinary projectors. Second, augmented reality (AR) applications using projection mapping technologies are introduced.*

### 1. INTRODUCTION

Projection mapping is one of the fundamental approaches to realize augmented reality (AR) systems. Compared to other approaches such as video/optical see-through AR, it can provide AR experiences without restricting users by head-mounted/use-worn/hand-held devices. In addition, the users can directly see the augmentations with natural field-of-view. On the other hand, there are several technical issues to be solved to display geometrically and photometrically correct images onto non-planar and textured surfaces. This invited talk introduces various techniques that our group proposed so far.

Leveraging the unique properties of projection mapping mentioned above, various interactive systems were proposed and investigated to support our daily activities, collaborative works, and machine manipulations. This talk also introduces such systems that were proposed by our group. Figure 1 shows the representative images of researches introduced in this talk.

### 2. FUNDAMENTAL TECHNOLOGIES

We developed fundamental technologies to overcome the technical limitations of current projection displays and to achieve robust user manipulation measurement under dynamic projection illuminations to realize flexible and robust interactive systems.

We use cameras, ranging from normal RGB cameras to near and far infrared (IR) cameras, to measure user manipulation as well as scene geometry and reflectance properties. For the user measurement, we apply IR cameras to robustly detect user's touch actions even under projection-based dynamic illumination. In particular, we propose two touch detection techniques; one measures finger nail color change by touch action using a near IR camera [22], while the other measures residual heat of user's touch on a surface using a far IR camera [9], [10]. For the scene measurement, we use projectors to project either spatial binary or uniform color patterns, and capture the projected results to measure scene depth [20] or reflectance properties. These geometric and photometric information is then used for radiometric compensation that neutralizes the texture of projection surface so that desired color is displayed on textured surfaces [4], [15], [24], [28].

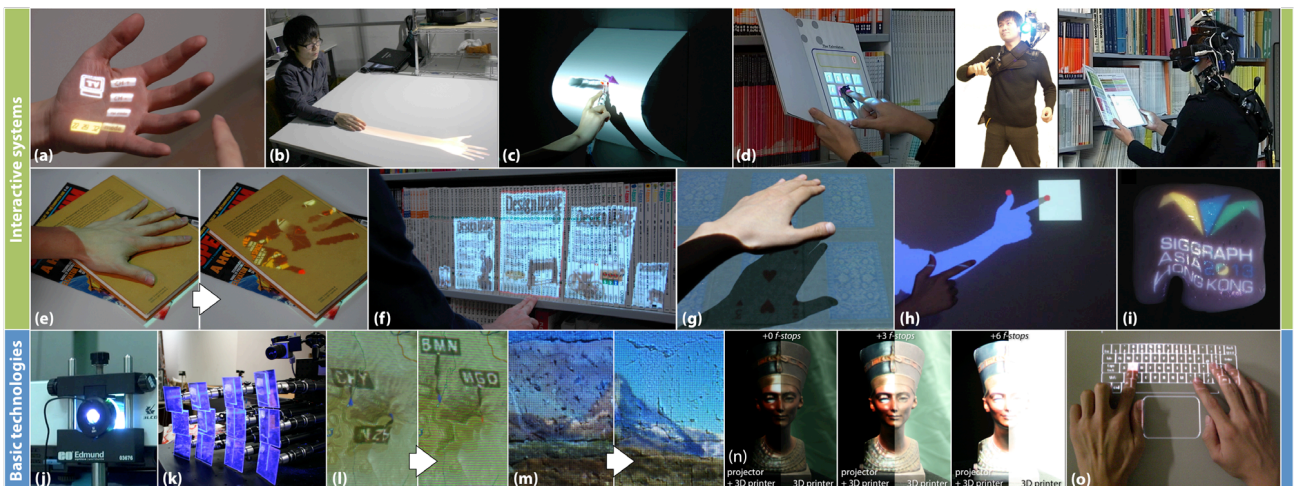

**Fig. 1** Representative images of researches introduced in this talk: (top and middle rows, clockwise) Palm interface [26], extended hand interface [17], shape deformation interface [27], wearable projection [13], deformation visualization [18], shadow pointing [25], shadow interface for browsing graphical information on a tabletop [5], and front cover projection [14] and see-through projection [10] for book search support. (bottom row, from left to right) Extended depth-of-field (EDOF) projector by fast focus sweep [7], synthetic aperture projection using a single projector and multiple mirrors for EDOF and shadow-less projection [16], view management of projected annotations [12], radiometric compensation [15], high dynamic range 3D projection mapping [21], and anywhere touch detection [22].

The followings are basic technologies on projection side, which are developed to change the appearance of real surfaces as freely as computer graphics do. In contrast, we realized high dynamic range projection systems by boosting the contrast of projection surface texture on relatively flat printed paper of E-ink display [2], on full color 3D printer output [21], and on surfaces with dynamic reflectance distributions using photochromic ink that allows us to modulate surface reflectance spatiotemporally [11], [23]. We also proposed to extend the DOF of projector by adopting fast focal sweep technique using electrically tunable liquid lens [7]. Extended DOF projection was realized with another approach, synthetic aperture projection, by which cast shadows could be removed as well [8], [16]. In order to realize wider field of view projection, we proposed pan-tilt projection and optimized the path of projected image to maximize the spatial resolution [6]. For displaying legible text information, we proposed an optimization method for projected annotation layout [12].

## 3. AR APPLICATIONS

We developed several interactive systems to demonstrate that AR by projection mapping technologies can support interactions between human and computer, human and human, and human and machine, respectively.

For human-computer interaction, we have investigated the following three topics: computing anywhere, shape design support, and book search support. Aiming at realizing the concept of computing anywhere, we developed wearable projection system in which a user can interact with graphical information projected onto nearby surfaces including her palm by touching it [13], [26]. To support shape design process, we proposed a system in which a user can manipulate the shape of a real surface, which is visually deformed by projected imagery [27]. The visualization of a real surface deformation by projection was also proposed [18]. We developed a system that can manipulate perceived haptic softness sensation by modifying the deformation visualization [19]. For book search support, we developed systems that allow users to search for their real books as they do for digital documents stored in their computers [10], [14].

For human-human communication, we developed an interactive tabletop system in which users can browse their private information by casting their shadows on the tabletop while sharing graphical information in non-shadow areas [5]. We also proposed shadow-based pointing interface for shared information displayed on a large screen around 10-feet away from users [25].

For human-machine interaction, we proposed an extended hand interface in which a user's hand is visually extended by projection so that she can manipulate distant appliances as she touch them [17]. While seating on a chair, the user can use the extended hand to tell a home service robot a specific distant object such as a cup that she wants the robot to bring or a specific place where the robot should clean up. We also proposed to enrich the face expression of an android robot by projecting high frequency face appearance information to realize more natural human-robot interaction [1]. In the extended hand research [17], we found an interesting phenomenon: a user felt the projected hand is her own hand. In another research, we found that warmth judgement of a touched object is affected by hand color, which is changed by projection [3]. These results indicate that we can alter human perception either by projecting user's body texture onto real surfaces or by projecting graphics onto a user's body.

## 4. CONCLUSION

This invited talk introduced projection mapping technologies for AR applications. In particular, it introduced the fundamental technologies to display desired images onto non-planar and textured surfaces. Then, it describes several interactive systems of projection-based AR. We are currently interested in applying projection mapping technologies to welfare and medical fields. For example, we are working on developing systems to support users suffering hand tremors for key input [29].

In future, we will focus more on investigating human perception issues too, which could allow us to develop more flexible and intuitive systems, with researchers in other domains such as cognitive science and neuroscience.

**ACKNOWLEDGEMENT**

This work was supported by JSPS KAKENHI Grant Number 15H05925.

**REFERENCES**

[1] A. Bermano, P. Brueschweiler, A. Grundhoefer, D. Iwai, B. Bickel, and M. Gross, "Augmenting physical avatars using projector based illumination," ACM Transactions on Graphics, Vol. 32, No. 6, pp. 189:1–189:10 (2013).

[2] O. Bimber and D. Iwai, "Superimposing dynamic range," ACM Transactions on Graphics, Vol. 27, No. 5, pp. 15:1–15:8 (2008).

[3] H.-N. Ho, D. Iwai, Y. Yoshikawa, J. Watanabe, and S. Nishida, "Combining colour and temperature: A

blue object is more likely to be judged as warm than a red object," Sci. Rep., Vol. 4, Article No. 5527 (2014).

[4] A. Grundhoefer and D. Iwai, "Robust, Error-Tolerant Photometric Projector Compensation," IEEE Transactions on Image Processing, Vol. 24, No. 12, pp. 5086-5099 (2015).

[5] M. Isogawa, D. Iwai, and K. Sato, "Making graphical information visible in real shadows on interactive tabletops," IEEE Trans. Vis. Comput. Graphics, Vol. 20, No. 9, pp. 1293–1302 (2014).

[6] D. Iwai, K. Kodama, and K. Sato, "Reducing motion blur artifact of foveal projection for dynamic focus-plus-context display," IEEE Trans. Circuits Syst. Video Technol., Vol. 25, No. 4, pp. 547-556 (2015).

[7] D. Iwai, S. Mihara, and K. Sato, "Extended depth-of-field projector by fast focal sweep projection," IEEE Trans. Vis. Comput. Graphics, Vol. 21, No. 4, pp. 462-470 (2015).

[8] D. Iwai, M. Nagase, and K. Sato, "Shadow removal of projected imagery by occluder shape measurement in a multiple overlapping projection system," Virtual Reality, Vol. 18, No. 4, pp. 245–254 (2014).

[9] D. Iwai and K. Sato, "Heat Sensation in Image Creation with Thermal Vision," In Proc. of ACE, pp. 213–216 (2005).

[10] D. Iwai and K. Sato, "Document search support by making physical documents transparent in projection-based mixed reality," Virtual Reality, Vol. 15, No. 2-3, pp. 147–160 (2010).

[11] D. Iwai, S. Takeda, N. Hino, and K. Sato, "Projection screen reflectance control for high contrast display using photochromic compounds and uv leds," Opt. Express, Vol. 22, No. 11, pp. 13492–13506 (2014).

[12] D. Iwai, T. Yabiki, and K. Sato, "View management of projected labels on non-planar and textured surfaces," IEEE Trans. Vis. Comput. Graphics, Vol. 19, No. 8, pp. 1415–1424 (2013).

[13] T. Karitsuka and K. Sato, "A wearable mixed reality with an on-board projector," In Proc. of IEEE ISMAR, p. 321 (2003).

[14] K. Matsushita, D. Iwai, and K. Sato, "Interactive bookshelf surface for in situ book searching and storing support," In Proc. of Augmented Human, pp. 2:1–2:8 (2011).

[15] S. Mihara, D. Iwai, and K. Sato, "Artifact reduction in radiometric compensation of projector-camera systems for steep reflectance variations," IEEE Trans. Circuits Syst. Video Technol., Vol. 24, No. 9, pp. 1631–1638 (2014).

[16] M. Nagase, D. Iwai, and K. Sato, "Dynamic defocus and occlusion compensation of projected imagery by model-based optimal projector selection in multi-projection environment," Virtual Reality, Vol. 15, No. 2, pp. 119– 132 (2011).

[17] S. Ogawa, K. Okahara, D. Iwai, and K. Sato, "A Reachable User Interface by the Graphically Extended Hand," In Proc. of IEEE GCCE, pp. 215–216 (2012).

[18] P. Punpongsanon, D. Iwai, and K. Sato, "Projection-based visualization of tangential deformation of nonrigid surface by deformation estimation using infrared texture," Virtual Reality, Vol. 19, No. 1, pp. 45-56 (2015).

[19] P. Punpongsanon, D. Iwai, and K. Sato, "SoftAR: Visually Manipulating Haptic Softness Perception in Spatial Augmented Reality," IEEE Transactions on Visualization and Computer Graphics, Vol. 21, No. 11, pp. 1279-1288 (2015).

[20] K. Sato and S. Inokuchi, "Range-Imaging System Utilizing Nematic Liquid Crystal Mask," In Proceedings of IEEE International Conference on Computer Vision (ICCV), pp. 657–661 (1987).

[21] S. Shimazu, D. Iwai, and K. Sato, "3d high dynamic range display system," In Proc. of IEEE ISMAR, pp. 235–236 (2011).

[22] N. Sugita, D. Iwai, and K. Sato, "Touch Sensing by Image Analysis of Fingernail," In Proc. of SICE, pp. 1520–1525 (2008).

[23] S. Takeda, D. Iwai, and K. Sato, "Inter-reflection Compensation of Immersive Projection Display by Spatio-Temporal Screen Reflectance Modulation," IEEE Transactions on Visualization and Computer Graphics, Vol. 22, No. 4, pp. 1424-1431 (2016).

[24] J. Tsukamoto, D. Iwai, and K. Kashima, "Radiometric Compensation for Cooperative Distributed Multi-Projection System through 2-DOF Distributed Control," IEEE Transactions on Visualization and Computer Graphics, Vol. 21, No. 11, pp. 1221-1229 (2015).

[25] H. Xu, D. Iwai, S. Hiura, and K. Sato, "User interface by virtual shadow projection," In Proc. of SICE-ICASE, pp. 4814–4817 (2006).

[26] G. Yamamoto and K. Sato, "PALMbit: A palm interface with projector-camera system," In Proc. of Ubicomp, pp. 276–279 (2007).

[27] K. Yamamoto, I. Kanaya, M. Hisada, and K. Sato, "The HYPERREAL design system – an mr-based shape design environment –," In Proc. of VSMM, pp. 201–208 (2005).

[28] T. Yoshida, C. Horii, and K. Sato, "A Virtual Color Reconstruction System for Real Heritage with Light Projection," In Proc. of VSMM, pp. 825–831 (2003).

[29] K. Wang, N. Takemura, D. Iwai, and K. Sato, "A Typing Assist System Considering Involuntary Hand Tremor," Transactions of the Virtual Reality Society of Japan, Vol. 21, No. 2, pp. 227-233 (2016).